\documentclass{article}

\usepackage{geometry}
\usepackage{natbib}
\usepackage{graphicx}

\title{Zombie Politics: Evolutionary Algorithms to Counteract the Spread of Negative Opinions}
\author{Ronald Hochreiter \and Christoph Waldhauser}
\date{April 2013}  

\begin{document}

\maketitle

\begin{abstract}
  This paper is about simulating the spread of opinions in a society
  and about finding ways to counteract that spread. To abstract away
  from potentially emotionally laden opinions, we instead simulate the
  spread of a zombie outbreak in a society. The virus causing this
  outbreak is different from traditional approaches: it not only
  causes a binary outcome (healthy vs infected) but rather a
  continuous outcome. To counteract the outbreak, a discrete number of infection-level
  specific treatments is available. This corresponds to acts of mild
  persuasion or the threats of legal action in the opinion spreading use
  case. This paper offers a genetic and a cultural algorithm that find the
  optimal mixture of treatments during the run of the simulation. They
  are assessed in a number of different scenarios. It is shown, that
  albeit far from being perfect, the cultural algorithm delivers
  superior performance at lower computational expense.
\end{abstract}

\noindent {\bf Keywords:} dynamic optimization, opinion propagation, epidemiology, evolutionary computing.

\section{Introduction}
The key aspect of marketing is to influence opinions. By conceiving
the political (in a democracy at least), as being yet another
application of marketing, politics become little more than
campaigning for votes, supporters and financial contributions.

When looking at influencing opinions, two use cases spring to mind:
(1) counteracting negative opinions online, and (2) to stop radical
political views from spreading to safeguard democracy.

With regard to the first use case, consider a Facebook page of a
company producing goods. A dissatisfied customer posts harsh criticism
to it. A common enough reaction is the threat of legal action against
that customer. If other users find this threat to be too harsh, they
will voice their opinions themselves, leading to a phenomenon similar
in nature to a more informal \textit{flamewar} but with much more dire
consequences for the involved stakeholders. Clearly, a more measured
response in these cases would help to stave off disaster. As
\citet{lin2010making} point out, this issue is becoming ever more
pressing as the availability of online opinion outlets increases.

In a less mundane example, counteracting radical opinions is equally
important to keep any democratic system well and functioning. Here,
radical opinions from the left and the right end of the political
spectrum threaten the very foundations of democracy. If these opinions
were to spread, the political system would crumble and fall. But how
to counteract these opinions? Clearly, using violent police crowd
control tactics to break up a group of peaceful protesters, will
overshoot and is likely to create sympathy for the group. Yet not
using police against militants will also lead to an undesired outcome.

Whenever treating political issues, research tends to become political
itself. To abstract away from the political, it is useful to consider
the spreading of opinions as being similar to the spreading of a
disease. This approach to modeling opinion dynamics has a sound basis
in the literature. For a direct application, see, for instance, \citet{adar2005tracking,
  sobkowicz2012opinion, lynch1998thought}, or see
\citet{moore2011extending} for feeding back into health sciences. While
diseases all have their medical idiosyncrasies, for the sake of
studying the \emph{spreading} of a disease, any disease will do. In order to
be able to also abstract away from medical realities, let there be a
fictional disease turning otherwise healthy humans into zombies. This
somewhat absurd approach to science has, however, been used quite
successfully in the past \citep{calderhead2010safe,
  munz2009zombies}. Their method is somewhat different, as it deals
with modeling the characteristics of infectious diseases with ordinary
differential equations. While this has its merits, differential
equations describing highly volatile, dynamic systems based on
stochastic selection quickly become unwieldy. Rather, we follow a
trajectory similar to \citet{crossley2011simzombie}, who modeled the
spread of zombies in a society using agent-based simulation.

We use this zombie approach with some modifications to model how a
disease (i.e.\@ radical opinions) spreads in society. The outbreak in
our model, however, can be counteracted using a selection of
treatments. Finding the right combinations of treatments is though
very far from being trivial. To this end, we offer two fine-tuned
evolutionary algorithms that will seek out the optimal combination of
treatments during the course of the outbreak.

The remainder of this paper is organized as follows. In the next
section, the disease model is introduced. After that we discuss the
evolutionary algorithms used for solving the curing puzzle in Section
\ref{sec:cure}. The algorithm is tested under varying underlying
assumptions. These results are then described in Section
\ref{sec:results} and discussed there after. There we follow along the
lines of \citet{cruz2011optimization} and present appreciations of the
algorithms' accuracy, stability and to a lesser extend
reactivity. Finally, we offer some conclusions.

\section{\label{sec:revil}Simulating opinion propagation}
As described above, when modeling the propagation of opinions, it
makes sense to assume a (fictional) disease. In the following we will
first describe the specific notions of opinions that need to be
considered when modeling them as a disease in a general way. We will
pay special attention on how and why opinion propagation differs from
classical epidemiological models. After that the employed disease
model is described in detail.

\subsection{Opinions as disease}
When modeling opinions as diseases, some notions apply that set these
models apart from classical epidemiological models on disease
spreading. Canonically \citep{teng1985comparison,sokolowski2011principles,brauer2011mathematical}, epidemiology employs models that identifies
distinct classes of healthy individuals that are susceptible to a
disease (S), the infected carriers of the disease (I) and individuals
that are immune to the disease either because of medical treatment,
seclusion or death (R). These models then assume that any individual
can potentially traverse these classes in turn:
\begin{equation}
S \to I \to R \to S
\end{equation}

Dependent on the classes included in the actual model, the model is
named in turn: the complete model is named SIRS, a model allowing only
for the move from healthy to infected with no healing or curing is
called SI.

Opinion propagation in the political process of a democracy sets
limits on the usability of these classes. A person that does not hold
an opinion can be classified as a member of class S, while convinced
advocates of that opinion will be in class I. Class R, those that are
removed from the population to halt the spreading of a disease or
particular opinion, has no real equivalent in any democratic
system. The locking-up of individuals holding any particular, perhaps
even iconoclastic opinion, is in essence incompatible with our
understanding of liberal democracy. Likewise, it is unlikely that any
political opinion will cause its holder's death. Therefore, opinion
propagation models for democracies must do without the R class and are
limited to be of type SI or perhaps SIS.\footnote{A noteworthy
  exemption are perhaps the Verbotsgesetze enacted in Austria and
  Germany under the impression of denazification. Here, indeed,
  prominent advocates of especially despicable opinions, are being
  held in prison and thus effectively removed from the population.}

Another factor that is crucial in the simulation of opinion
propagation in human populations is the topology of the underlying
network. It has been shown, that human interaction is organized in the
fashion of small-world networks \citep{rodriguez2012decision,wessels1999system,eubank2004modelling,amaral2000classes}. These networks govern how individuals
are linked to each other. Most prominently, the small world property
states that most individuals are connected to most other individuals
via only a hand full of other individuals. This requires some
individuals to function as hubs, that are connecting many people.

These small world networks are on one hand quite fault tolerant and
avoid the breaking down of communication if single individuals are
removed. More importantly for the case at hand, however, is that they
also facilitate the spreading of disease, once central nodes that
connect many individuals are infected. 

Even though these models are representative of the way humans
interact, they add an additional level of complexity to the modeling
of disease and opinion propagation. Therefore, the simplest models of
disease propagation focus on a relaxed model that ignores these
network properties. Rather, the simpler model assumes a population
that is well-mixed. Here, every individual is connected to every other
with the same probability. This boils down to a mean-fields model in
the terminology of physics' parlance.

\subsection{Disease model}
As detailed above and elsewhere \citep{rahmandad2008heterogeneity,patlolla2006agent}, there are compelling reasons to diverge from
canonical epidemiological models when dealing with opinions. We
therefore employ a SI-model ignorant to complex network topologies. In
our model of opinion propagation, the infected are not being secluded
from the population, and every individual has an equal, mean
connectedness to every other individual. However, further
modifications apply that we will describe in the following.

Here we use the Revil zombie outbreak simulator
\citep{waldhauser2013revil}\footnote{All computations were done in R
  \citep{Rcore2012}; plots were produced using \emph{ggplot2} \citep{ggplot2}.}. This
package allows to model not only the conventional dichotomous disease
outcome of healthy versus infected. Rather, it assumes a continuous
spectrum of infectiousness. Humans carry parts of the disease within
them, at all times. But only once they pass a certain threshold, they
will turn into zombies and start infecting others.

This notion of a continuously conceptualized disease, aptly named
C-Virus, is quite similar to the way opinions work. Take for example
xenophobia. While only very convinced xenophobes will start
propagating these views, who can honestly say to not have reservations
against the unknown? Or while many cannot agree with every point Marx
ever made, who would not cherish the overall improvement of labor
conditions? Aspects of even the most radical of views are present in
everyone of us, to a varying degree.

In its design, the disease model is quite simple. A society of a given
size is initialized to a random state of infection. This is another
point of departure from the traditional model that always assumes a
single\footnote{or a small number of} patient zero. But who could ever
identify e.g.\@ the first xenophobe?  In addition, there is a number
of zombies, i.e.\@ humans that carry so much of the disease within
them, that they can corrupt others. From this initial state forward,
interaction between humans and zombies follows a stochastic
trajectory. Humans that come in contact with zombies have a certain
probability of having their infectiousness increased. Once a human
crosses a certain threshold, the disease becomes acute and that human
turns into a zombie, lusting to attack others. If unchecked, the
disease will continue its spread until all humans have been turned
into zombies.

This has serious ramifications for the disease model. While
conventional models assume distinct classes of healthy humans and
infected. In our model, these distinctions blur. In the employed fuzzy
model of disease it is difficult to draw a line, as even healthy
individuals carry the disease to a certain extent. If considering the
threshold described above as the discriminatory boundary, all those not
yet carrying the disease to an extent that crosses the threshold, are
to be considered healthy; the remainder, i.e.\@ zombie population has
to be counted as infected. 

While in canonical zombie lore, there is seldom a cure to the
outbreak, the C-Virus can be treated. However, there is a caveat. As
in real life opinions, treatments are specific to the seriousness of
the infection: a mildly infected individual needs different
medications than a raging wild zombie. To simplify the model, we
assume that there is a limited set of discrete treatments that form a
cure that can be administered to the general public. Each treatment is
only effective for a certain range of infected. In the cure mixture,
we need to specify the dose of each treatment, i.e.\@ the probability
of that treatment being used on someone.

Finding a cure, that over many days manages to keep the zombies at
bay, proved rather difficult. The following section describes how
evolutionary algorithms help in finding that optimal mixture.


\section{Searching for a cure}
\label{sec:cure}
When looking at societies ravaged by our virus, two key indicators can
be analyzed. One is the mean infection rate ($mir$), which is the
average degree of infectedness in the entire society. A society
composed entirely of healthy humans would exhibit $mir=0$ while in one
where humans had died out and all zombies be fully infected, the maximum of $mir=1$ would be reached. Any
cure for the C-virus is more successful, if the mean infection rate is
kept lower than under application of a different cure.

When it is not possible to effectively root out the disease, it might
still be possible to slow its progression. The second quantity of
interest is thus the last day a human was encountered in the society
($lHd$). A cure is successful when it manages to postpone the advent
of the zombie apocalypse, i.e.\@ the extinction of healthy humans.

For any disease, one of the key variables is virulence ($v$). That is,
according to the medical subject headings database MeSH
\citep{rogers63}, the ability of a pathogen to cause disease in its
host. In a simplified model, one can think of virulence as the
probability of becoming ill after being in contact with the virus.

In the following, two scenarios were assumed: one where virulence was
assumed to be relatively low, and one where virulence was set at a
dramatically high level, where virtually every human coming in contact
with a zombie is being infected automatically.

Both scenarios were computed using different search strategies. Each
computation was repeated for a number of  times and the mean
and standard deviation computed. Additionally, a baseline was
established using a simulation where no cure at all was available.

The search strategies employed are (a) a purely genetic algorithm and
(b) a cultural algorithm, as proposed by
\citet{reynolds1994introduction}. A detailed description of the used
algorithms is given in the following subsections.

When searching for an optimal cure, i.e.\@ a mixture of treatments,
the algorithm only has a limited set of generations available, before
that cure needs to be applied to society. This allowance of
generations per day will prove to have a dramatic effect on the
algorithms' ability of finding effective cures early on during the
simulation. After application of the cure, the simulation is advanced
a day forward, with new infections taking place. This turns the
optimization problem into a dynamic version of the knapsack problem
\citep{cruz2011optimization,branke2002evolutionary}. While in the traditional knapsack
problem, an optimal combination of items fitting into a static
knapsack is the target of the optimization, here, the features of the
knapsack change while the search is still well under way. To properly
identify this kind of optimization problem, we propose the name of
dynamic continuous disease treatment problem.

\subsection{Genetic search}
The genetic algorithm seeking out the optimal mixture of treatments
follows along traditional lines of genetic computing. The set of ten
treatment intensities making up the cure comprises the chromosome,
with each treatment being a real-valued gene in the domain of $(0, 1)$.

The goal of the optimization was to find a cure that minimizes the
society's mean infection rate if applied. Fitness was thus established
by simulating the application of each chromosome's encoded solution to
the society at the current state, simulating the progression of the
disease for one day, and then estimate the resulting mean infection rate. A
chromosome was found to be the fitter the lower the caused mean
infection rate was.

The genetic operators implemented are cross-over replication and
mutation. Cross-over computes the mean between all gene pairs and
returns it as the child's value. Mutation randomly replaces a gene
with a uniformly distributed value. The best solution clones itself to
guarantee monotonicity. Breeding is organized in a tournament
selection scheme, with tournament size $k=5$. The winner of each
tournament enters a breeding pool, from which pairs are selected at
random to create offspring. 


\subsection{Cultural search}
One of the greatest benefits of cultural algorithms is that they can
quicken the search for an optimum by retaining a history of previously
tried solutions (situational knowledge) and including search-related meta-information (normative knowledge) \citep{reynolds2008computing}. While in
the zombie infection scenario, meta information as in its real world
applications is not readily available, the retaining of best solutions was
included into the algorithm.

In the traditional purely genetic approach, two extraordinarily fit
solutions are selected at random to create offspring. The cultural
algorithm changes this behavior drastically. After a generation has
been evaluated, the fittest solution writes itself into a
meta-storage, called belief space. This belief space then is used to
influence all of the offspring created. Influence is implemented by
replacing either mother or father of a child solution at random with
the information stored in the belief system. Other than the belief
space influencing mating partner selection, the algorithm remained
unchanged from its genetic brother. 

The details of both algorithms' parametrization can be found in Table
\ref{tab:params}.

\begin{table}
  \centering
  \caption{\label{tab:params} Parametrization of the algorithms.}
  \begin{tabular}{rr}
    \hline
    Quantity&Value\\
    \hline
    Population size & 100\\
    Tournament size & 5\\
    Mutation probability &0.05\\
    \hline
  \end{tabular}
\end{table}


\section{Results}
\label{sec:results}
To demonstrate the suitability of the evolutionary approaches
described above for finding cures to C-Virus outbreaks, a number of
scenarios was considered. Pertaining to the real world applications,
different degrees of virulence were assumed: a rather mild disease
with a low virulence of $v=0.3$ and an extremely aggressive one
($v=1$), where every person meeting a zombie becomes
infected. Obviously, the former scenario leaves the algorithm more
time to react to the outbreak, while the latter requires swift action
to effectively combat the threat. All simulation parameters are given
in Table \ref{tab:simparams}.

\begin{table}
  \centering
  \caption{\label{tab:simparams}Simulation parameters. \textbf{Bold} parameters were varied between scenarios.}
  \begin{tabular}{rr}
    \hline
    Parameter&Value\\
    \hline
    Infection threshold & 0.75\\
    Number of treatments&10\\
    Range of treatment effectiveness&$\pm 10\%$\\
    Number of replications&10\\
    \textbf{Virulence settings}&0.3, 1\\
    \textbf{Generations/day optimization allowance}&1, 5, 25\\
    \hline
  \end{tabular}
\end{table}

The second parameter varied was the time the algorithms got to
optimize before they were forced to apply their respective optimal
solutions to the society and the simulation advanced a day
forward. This generations per day allowance, $gd$, was simulated at
settings of 1, 5 and 25. A value of 50 was tried as well, but omitted
from later analysis, as the differences to $gd=25$ proved negligible.

All scenarios started from the same society of 50 individuals. The
society was initialized with random infection values, ranging for
humans from $0$ to $0.35$ and for zombies from $0.8$ to $1$. The
proportion of zombies in this initial society was set at $0.1$. This
resulted in a society mean infection rate of $0.25$ at the onset of
the simulation.

Because the progression of the disease is stochastic in nature, all
simulations were replicated 10 times. All reported figures are
averages over these replications, with the standard deviation given,
when appropriate.

The untreated disease with low virulence will have eradicated the last
human after 160
($\sigma =9.4$) days. For a highly
virulent strand of the disease, 71
($\sigma =3.4$) days will be enough to
leave nothing but babbling zombies. The mean infection rates for both
scenarios were established at 0.59
($\sigma =0.03$) and
0.81
($\sigma =0.01$), respectively.

\begin{figure*}
  \centering
  \includegraphics{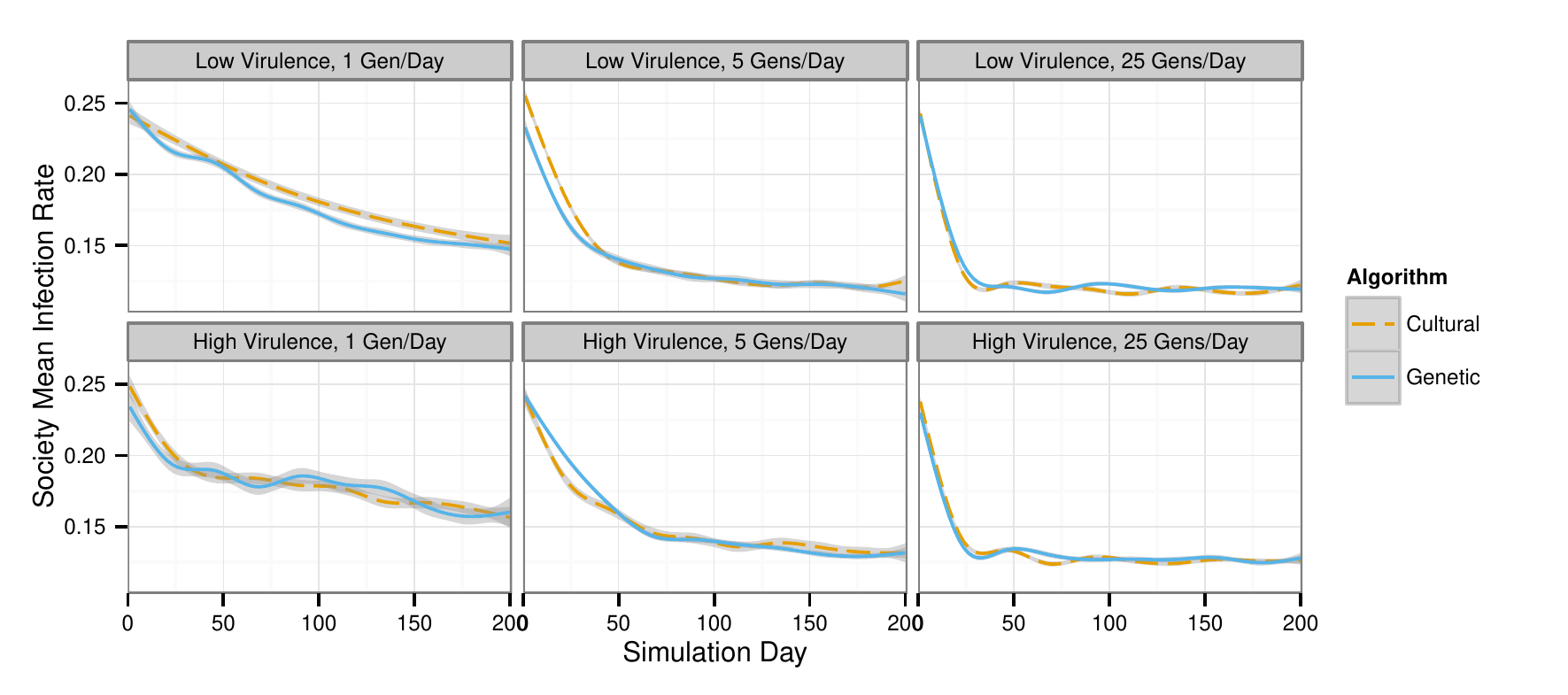}
  \caption{\label{fig:mirs}Algorithms' performances in low and high virulence settings, with different generations per day optimization allowances. GAM estimates of 10-fold bootstrapped results, shaded areas indicate 95\% confidence bands.}
\end{figure*}

Fortunately for humankind, the algorithms were able to counter the
zombie threat successfully. Figure \ref{fig:mirs} compares the
effectiveness of both algorithms in all six scenarios. As the scenario
assuming a highly virulent disease produces more pronounced
differences and for the sake of brevity, only the results from that
scenario will be presented in the following.

The most effective cure translates directly into the lowest mean
infection rate achieved during the course of the simulation. Both
algorithms' achievements in that regard are given in Table
\ref{tab:mirs-hi}. 

\begin{table}[ht]
\centering
\caption{Lowest MIR achieved (high virulence)} 
\label{tab:mirs-hi}
\begin{tabular}{rrrrr}
  \hline
GA Mean & GA SD & CA Mean & CA SD & Generation \\ 
  \hline
0.16 & 0.03 & 0.16 & 0.02 &   1 \\ 
  0.13 & 0.00 & 0.13 & 0.01 &   5 \\ 
  0.13 & 0.01 & 0.13 & 0.01 &  25 \\ 
   \hline
\end{tabular}
\end{table}

Besides finding good solutions, also the reliability of solutions
produced at any given stage of the simulation are of importance. To assess this stability, the variance of the ten replications for each day was computed. The
results across algorithm type and generations per day optimization
allowance are presented in Table \ref{tab:mirspread-hi}. The lower the
variation is, the more stable the results are.
\begin{table}[ht]
\centering
\caption{Stability of achieved MIRs (high virulence)} 
\label{tab:mirspread-hi}
\begin{tabular}{rrr}
  \hline
Genetic & Cultural & Generation \\ 
  \hline
0.03 & 0.03 &   1 \\ 
  0.01 & 0.02 &   5 \\ 
  0.01 & 0.01 &  25 \\ 
   \hline
\end{tabular}
\end{table}

Finally, when combating potentially dangerous political attitudes
(and zombie inducing diseases), it is crucial to score good results
early on. Table \ref{tab:fma-hi} gives the days it took, for the 
algorithm's solutions to enter a 5\%-neighborhood of their final
results. The earlier an algorithm enters this neighborhood, the more
suited it will be to combat even highly virulent strands of the
disease.

\begin{table}[ht]
\centering
\caption{First day on which algorithm entered 5-percent neighborhood of final result (high virulence)} 
\label{tab:fma-hi}
\begin{tabular}{rrr}
  \hline
Genetic & Cultural & Generation \\ 
  \hline
141 & 139 &   1 \\ 
  122 &  90 &   5 \\ 
   28 &  28 &  25 \\ 
   \hline
\end{tabular}
\end{table}

\begin{figure*}
  \centering
  \includegraphics{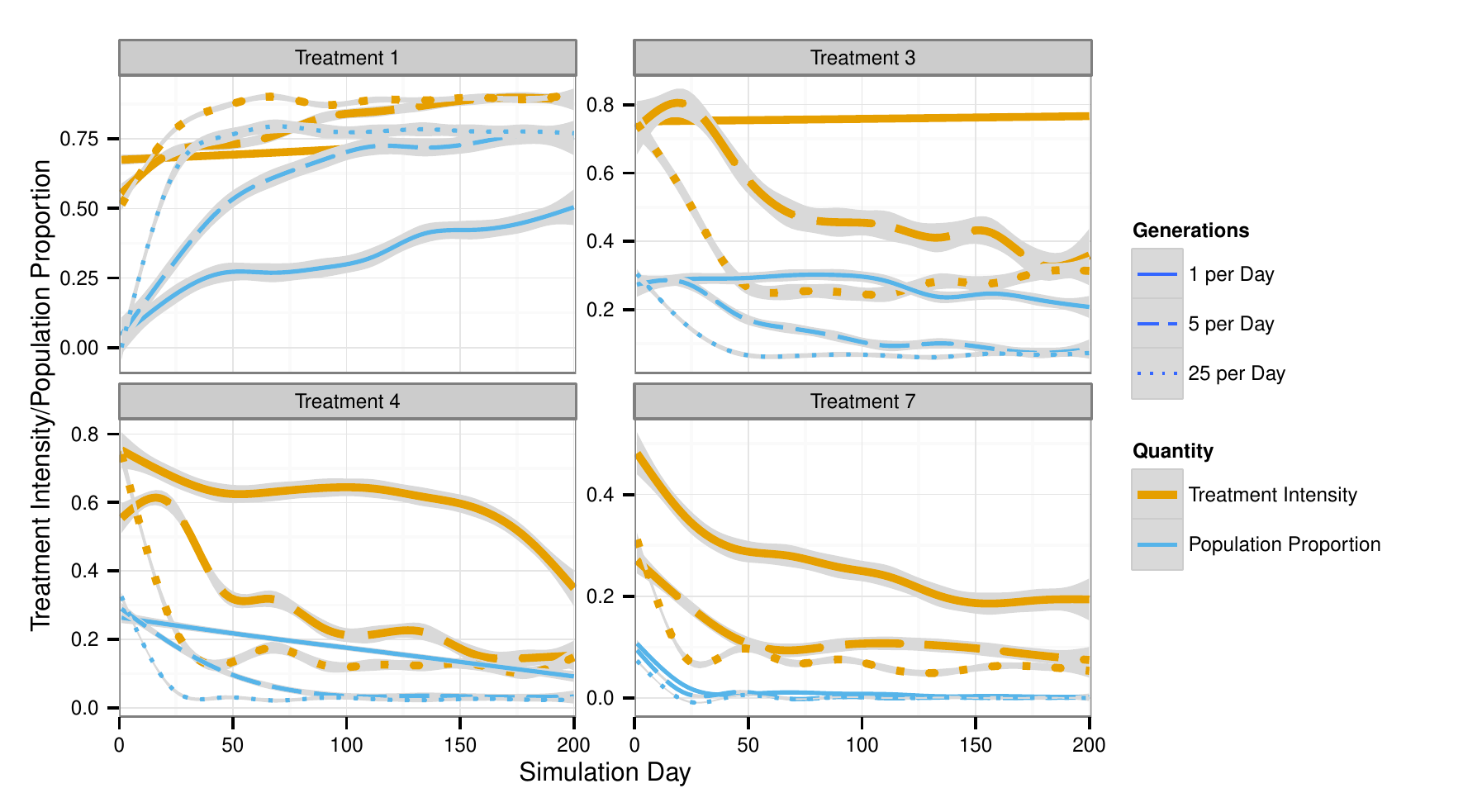}
  \caption{\label{fig:treatments}Selected treatment intensities as produced by cultural search algorithms with different generations per day optimization allowances in comparison with true proportion of infected in society. GAM estimates of 10-fold bootstrapped results, shaded areas indicate 95\% confidence bands.}
\end{figure*}

When turning to the solutions produced, a dynamic optimization problem not only warrants an examination of the final results. Rather, it is mandated to analyze the development of the solutions during the course of the simulation. Figure \ref{fig:treatments} displays the development of four treatments chosen to exemplify the overall development. Each panel depicts a single treatment. The lighter lines describe the intensity of the respective treatments as it was applied over the course of the simulation. The darker lines show how large the proportion of the population that was positively affected by that treatment actually was. Note, that these figures was unknown to the optimization algorithms, as would be the case during real world applications. 

In this section we described the results of both genetic and cultural
algorithms in solving the dynamic disease treatment problem. A crucial
parameter proved to be the numbers of generations granted to an
algorithm before it has to commit its solution, and the simulation is
advanced forward one day. In the following section we will discuss these
results and their implications.

\section{Discussion}
\label{sec:discussion}

The optimization's overall goal was to reduce the mean infection rate
in the society. Translating the mean infection rate into the realm of
real (political) world applications, this means to reduce the overall
(negative) attitudes held in a society regarding a specific topic. As
demonstrated by the baseline escalation, the spreading of these bad
views knows no bounds, if unchecked. Fortunately, both algorithms even
with low generations per day optimization allowances were able to keep
a balance. Indeed, in this regard, both algorithms proved to be quite
indistinguishable. Obviously, the more computing time (in terms of
generations per day) an algorithm was granted, the further down the
mean infection rate could be pushed. It is also noteworthy, that no
algorithm managed to eradicate the disease entirely.

Similarly, both algorithms were on par regarding the stability of
their obtained solutions, exhibiting only minuscule differences. If
looking at these figures from the point of view, that containing the
outbreak is the ultimate goal, these results are encouraging. However,
if one wants to treat the disease for good, both algorithms are
disappointing. As high stability also means it is unlikely to produce
even better results, hopes for curing, i.e.\@ rooting out the disease, are
lost.

The perhaps most interesting feature is the time it took to achieve
nearly optimal solutions. Here, cultural algorithms truly shine. While
the brutal variant, where algorithms are barely allowed to optimize
and need to apply their solutions almost instantly, shows little
difference, it is the 5 generations per day allowance that exhibits
radical differences. Here, the cultural algorithm finds optimal
solutions more than a month ahead of its purely genetic
brother. Only when providing both algorithms with 25 generations per
day of optimization allowance, the differences subside.

This is a crucial observation. Allowing an algorithm to spend that
much of computing power is expensive both in the time and money
dimensions. Obviously, the detour via a belief space induced culture is
worth its price: despite adding complexity to the optimization, it
results in better outcomes, early on. In the application of combating
negative opinions, these days won by more clever optimization,
translate directly into being able to more quickly counteracting views
that cost real market share or, more dramatically, democratic quality of a
society.

\section{Conclusions}
\label{sec:conclusions}
In this paper, we aimed at examining the ways of counteracting the
spread of negative opinions. To abstract away from the harsh realities
of politics and marketing, we used the analogy of a hypothetical
zombie-state inducing virus spreading in an artificial society. The
disease model diverged significantly from canonical epidemiological
literature. By conceiving the disease as a continuous outcome variable,
the framework of diseases has been made accessible to the propagation
of opinions. The omission of a class of deceased or otherwise immune
to an opinion individuals was mandated. The focus on a simple
population structure forgoing complex network topologies is adequate
for preliminary research.

As in real life, the goal was to contain the outbreak and thus reduce
the overall amount of negative views held within the society. This
amounts to a dynamic optimization problem which used to be very hard
to solve in the past. By using two different evolutionary approaches,
we were able to contain, but not necessarily eradicate the
outbreak. Clearly, there remains room for improvement here. If, after
simulation day 200, we would have allowed the simulation to continue
without the application of constantly updated treatment solutions, the
disease would have sprung up again and eventually taken over society
at large.

However, the simulation at hand has demonstrated that making the right
choices in counteracting negative opinions does matter. Indeed,
countering opposite attitudes with maximum (political) force is bound
to produce negative consequences. Rather, the situation needs to be
analyzed and judged dispassionately. Only then can solutions be found,
that keep the opposing force's proponents at bay. Using evolutionary
algorithms shows but one way of achieving this goal.

The complexity added by using cultural algorithms proved to be well
invested in producing solutions that were able to contain the outbreak
more quickly and at less computational expense than the purely genetic
variant. Thus, we are left with two areas that mandate improvement: (1)
reduction in computational demand, and (2) providing us with the
ability to effectively eradicate the threat of disease now and always.

We would like to point out to a number of issues that would warrant
further research. Chiefly, this would concern investigations in
evolutionary algorithms that succeed in eradicating the disease for
good. Further, applying the simulation to the more prevalent class of
complex networks would prove worthwhile. Finally, relaxing the models
assumptions, perhaps even (re-)introducing the class of deceased would
prove especially intriguing.

The results clearly shows that we are in need of an evolutionary
algorithm that has the power to deliver solutions that are able to
eradicate the disease once and for all, using only an acceptable
amount of computational power. While \citet{hochreiter13} and
\citet{guo2011novel} both seem to point in the right direction, whether
their algorithms are up to the challenge remains to be seen.

\nocite{miller1999policy}

\bibliographystyle{plainnat}
\bibliography{main}

\end{document}